\documentclass{article}
\usepackage{emulateapj,pstricks,apjfonts}
\def\chandra	{{\em Chandra}\/}
\def\xmm	{{\em XMM}\/}
\def\asca       {{\em ASCA}\/}
\def\astroe     {{\em Astro-E}\/}
\def\einstein   {{\em Einstein}\/}
\def\rosat      {{\em ROSAT}\/}
\def\hfifty     {$H_0$=50~km$\;$s$^{-1}\,$Mpc$^{-1}$}
\def\gax	{\gtrsim}

\def\msun	{~$M_{\odot}$}
\def\msunyr     {$M_{\odot}\;$yr$^{-1}$}
\def\am         {$^{\prime}$}
\def\as         {$^{\prime\prime}$}
\def\kms        {~km$\;$s$^{-1}$}
\def\cmcube     {~cm$^{-3}$}
\begin{document}

\submitted{ApJ in press; astro-ph/9812005 v2}

\lefthead{CYGNUS A, A3667, AND A2065} 
\righthead{MARKEVITCH, SARAZIN, \& VIKHLININ}

\title{PHYSICS OF THE MERGING CLUSTERS CYGNUS A, A3667, AND A2065}

\author{Maxim Markevitch\altaffilmark{1,2}, Craig L. Sarazin\altaffilmark{3}, 
and Alexey Vikhlinin\altaffilmark{1,2}}

\altaffiltext{1}{Harvard-Smithsonian Center for Astrophysics, 60 Garden St.,
Cambridge, MA 02138; maxim, alexey @head-cfa.harvard.edu}

\altaffiltext{2}{Space Research Institute, Russian Academy of Sciences}

\altaffiltext{3}{Astronomy Department, University of Virginia,
Charlottesville, VA 22903; cls7i@virginia.edu}

\begin{abstract}

We present \asca\ gas temperature maps of the nearby merging galaxy clusters
Cygnus A, A3667, and A2065. Cygnus A appears to have a particularly simple
merger geometry that allows an estimate of the subcluster collision velocity
from the observed temperature variations. We estimate it to be $\sim
2000$\kms. Interestingly, this is similar to the free-fall velocity that the
two Cygnus A subclusters should have achieved at the observed separation,
suggesting that merger has been effective in dissipating the kinetic energy
of gas halos into thermal energy, without channeling its major fraction
elsewhere (e.g., into turbulence). In A3667, we may be observing a spatial
lag between the shock front seen in the X-ray image and the corresponding
rise of the electron temperature. A lag of the order of hundreds of
kiloparsecs is possible due to the combination of thermal conduction and a
finite electron-ion equilibration time. Forthcoming better spatial
resolution data will allow a direct measurement of these phenomena in the
cluster gas using such lags. A2065 has gas density peaks coincident with two
central galaxies. A merger with the collision velocity estimated from the
temperature map should have swept away such peaks if the subcluster total
mass distributions had flat cores in the centers. The fact that the peaks
have survived (or quickly reemerged) suggests that the gravitational
potential also is strongly peaked. Finally, the observed specific entropy
variations in A3667 and Cygnus A indicate that energy injection from a
single major merger may be of the order of the full thermal energy of the
gas. We hope that these order of magnitude estimates will encourage further
work on hydrodynamic simulations, as well as more quantitative
representation of the simulation results, in anticipation of the \chandra\
and \xmm\ data.

\end{abstract}


\section{INTRODUCTION}

Mergers of galaxy clusters are the most energetic events in the Universe
since the Big Bang, with the total kinetic energy of the two colliding
subclusters reaching $10^{63-64}$ ergs. In the course of a merger, a major
portion of this energy, that carried by the subcluster gaseous halos,
dissipates in the intracluster gas, heating it and possibly generating
turbulence, magnetic fields, and relativistic particles.  It is interesting
to determine how much of this energy goes into turbulence and elsewhere,
because this bears on the cosmologically important issue of the accuracy of
the cluster total masses (e.g., Loeb \& Mao 1994). A merger in a cooling
flow cluster also subjects the cooling flow to ram pressure (e.g., McGlynn
\& Fabian 1984; Fabian \& Daines 1991) and other destructive effects,
directly probing the gas pressure and the depth of the gravitational
potential in the cluster center. Detailed maps of mergers may also provide
information on thermal conduction (and therefore magnetic fields) and
electron-ion equilibration timescale in the intracluster plasma.

Merger shocks should produce strong spatial variations of the intracluster
gas temperature and density (for a simulations review see, e.g., Burns
1998), such as those observed by \asca\ and \rosat\ in several dynamically
active clusters (Henry \& Briel 1995; Henriksen \& Markevitch 1996; Honda et
al.\ 1996; Donnelly et al.\ 1998, 1999; Churazov et al.\ 1999; Markevitch et
al.\ 1994, 1996, 1998 [hereafter MFSV]). In this paper, we present the first
temperature maps for three particularly interesting nearby merging clusters,
Cygnus A, A3667, and A2065, obtained with \asca. Despite the modest spatial
resolution of these maps, some qualitative conclusions regarding the physics
of mergers and the shape of the cluster gravitational potential can be made.
Unlike previous quantitative merger studies (e.g., Roettiger et al.\ 1998,
1999), we have chosen not to use the help of hydrodynamic simulations for
the data interpretation, in order to examine what can be learned from a
purely observational point of view. We use \hfifty.

\section{\asca\ TEMPERATURE MAPS}

\asca\ (Tanaka et al.\ 1994) is capable of mapping the gas
temperature in nearby clusters. MFSV analyzed Cygnus A, A2065 and A3667 as
part of a larger \asca\ cluster sample and presented average temperatures
and radial temperature profiles for these clusters.  The derivation of the
projected gas temperature maps is described in detail in that paper and
references therein. It involves the convolution of a two-dimensional
temperature and emission measure model with the \asca\ response, and fitting
the temperatures in a number of spatial regions. As models for the emission
measure, we used \rosat\ PSPC images, taking into account the detected
temperature variations.  There is no PSPC image for A2065, so \einstein\ IPC
and \rosat\ HRI images were combined. The Cygnus A AGN spectrum was fitted
simultaneously with all the gas temperatures as described in MFSV. The
resulting gas temperature maps are shown in Figs.\ 1 and 3({\em b,c}). Error
estimates include all known calibration uncertainties. For the
better-resolved Cygnus A and A3667, we also show maps of the gas specific
(per particle) entropy, defined as $\Delta s\equiv s-s_0=\frac{3}{2}k\;{\rm
ln}\left[ (T/T_0)(\rho/\rho_0)^{-2/3}\right]$, where the subscript 0 refers
to any fiducial region in the cluster. For a qualitative estimate, we
approximate $\rho/\rho_0 \sim (S_x/S_{x0})^{1/2}$, where $S_x$ is the
cluster X-ray surface brightness in a given region.

All three clusters display significant gas temperature variations which,
together with their complex X-ray brightness morphology, indicate ongoing
mergers. Below we discuss some interesting implications for each cluster.

\section{Cygnus A}
\label{sec:cyga}

From the temperature and brightness maps, Cygnus A ($z=0.057$) appears to
have a particularly simple merger geometry, with two similar $T\simeq 4-5$
keV subclusters colliding head-on and developing a shock between them.
Substructure is also observed in the optical (Owen et al.\ 1997). Taking
advantage of this simplicity, we try to estimate the subcluster collision
velocity directly from the observed gas temperature variations.

For such an estimate, we assume that the region between the Cygnus A
subclusters can be approximated by a one-dimensional shock. Then following
Landau \& Lifshitz (1959) and using the Rankine--Hugoniot jump conditions,
one can derive the difference of the gas flow velocities before and after
the shock as a function of the respective gas temperatures:
\begin{equation}
u_0-u_1 = 
\left[\frac{k T_0}{\mu m_p}(1-x)\left(\frac{1}{x} \frac{T_1}{T_0}-1\right)
\right]^{1/2},
\end{equation}
where velocities are relative to the shock surface, indices 0 and 1 denote
quantities before and after the shock (\mbox{$0<u_1<u_0$} and $T_0<T_1$),
$\mu=0.6$ is the plasma mean molecular weight, and
\begin{equation}
x\equiv \frac{u_1}{u_0}= 
\left[\frac{1}{4}\left(\frac{\gamma+1}{\gamma-1}\right)^2
\left(\frac{T_1}{T_0}-1\right)^2 +\frac{T_1}{T_0}\right]^{1/2} 
-\frac{1}{2}\frac{\gamma+1}{\gamma-1}\left(\frac{T_1}{T_0}-1\right).
\end{equation}
The quantity $x$ is the inverse of the shock compression, $x = ( \rho_1 /
\rho_0 )^{-1}$, and may also be determined from the increase in X-ray
surface brightness at the shock as sometimes seen in X-ray images (e.g.,
Fig.\ 2 below). The application of these equations to the X-ray spectra
assumes that the gas within the shocked region is nearly isothermal, and
that electrons and ions reach equipartition. We will touch on these issues
below.

For a symmetric merger in Cygnus A, we assume that the shocked gas on
average is at rest relative to the center of mass. The collision velocity is
then $\Delta u_{\rm cl}=2(u_0-u_1)$. The pre-shock temperature can be
estimated from the symmetric regions on the opposite sides of the
subclusters, $T_0\approx 4\pm 1$ keV, and the post-shock value is
$T_1\approx 8^{+2}_{-1}$ keV (90\% errors). The X-ray surface brightness in
these regions is roughly consistent with the compression expected in the
shock.  Substituting these temperatures, we obtain $\Delta u_{\rm cl}\approx
2200^{+700}_{-500}$ km/s.

Obviously, this simplistic shock model ignores many important details such
as gas density gradients in subclusters, velocity and temperature gradients,
and outflows along the shock plane. These effects are best taken into
account by hydrodynamic simulations, as in, e.g., Roettiger et al.\ (1998,
1999). To gauge the importance of these effects, Ricker \& Sarazin (1999)
have simulated head-on cluster mergers similar to the one in Cygnus A. They
find that there are significant temperature and velocity gradients within
the shocked region which result from adiabatic compression, yet the simple
shock velocity argument given above is accurate to about 20\%.

Interestingly, the collision velocity found above is close to the free-fall
velocity of $\sim$2000\kms\ that two similar 4--5 keV, $R\approx 1$ Mpc
clusters should have achieved had they fallen from a large distance to the
observed separation of 1 Mpc. This consistency may suggest that kinetic
energy of the gas in colliding subclusters is effectively thermalized, since
otherwise (e.g., if a major fraction of this energy were channeled into
turbulence, magnetic fields, or cosmic rays) the shock temperature would not
be as high as observed.%
\footnote{Note, however, that such a consistency by itself cannot
prove that in general there are no dynamically important turbulence or
magnetic fields in clusters, since for the free-fall velocity calculation,
masses are estimated from the gas temperature assuming the absence of such
effects.}

\section{A3667}
\label{sec:3667}

A3667 is a spectacular merger at $z=0.053$. It also has an extensive double
radio halo outside the central region (R\"ottgering et al.\ 1997),
straddling our region of measured temperatures as shown in Fig.\ 1.  The
galaxy distribution is bimodal (Proust et al.\ 1988), with the main
component around the cD galaxy in region 5 of Fig.\ 1 and the secondary
component around the second-brightest galaxy in region 14. The hottest gas
is observed between these subclusters, indicating that they are presently
colliding (although not necessarily for the first time). The brightness peak
seems to lag behind the cD galaxy (Fig.\ 1), consistent with this picture.
The merger is apparently more complex than in Cygnus-A, but for order of
magnitude estimates, we still apply the simple modeling of \S\ref{sec:cyga}.
For pre- and post-shock temperatures of 4 keV and 10 keV, the collision
velocity is likely to be of the order of $2(u_0-u_1)\sim 2500-3000$\kms, and
the velocities of the two shocks propagating from the site of the collision
in region 10 of the map are $u_1\sim 1000$\kms\ relative to the collision
site.

%%%%%%%%%%%%%%%%%%%%%%%%%%%%%%%%%%%%%%%%%%%%%%%%%%%%%%%%%%%%%%%%%%%%%
\pspicture(0,0.4)(8.8,9.9)
%\psgrid(0,0)(9,10)

\rput[tl]{0}(-0.5,9.7){\epsfxsize=8.8cm
\epsffile[25 185 550 588]{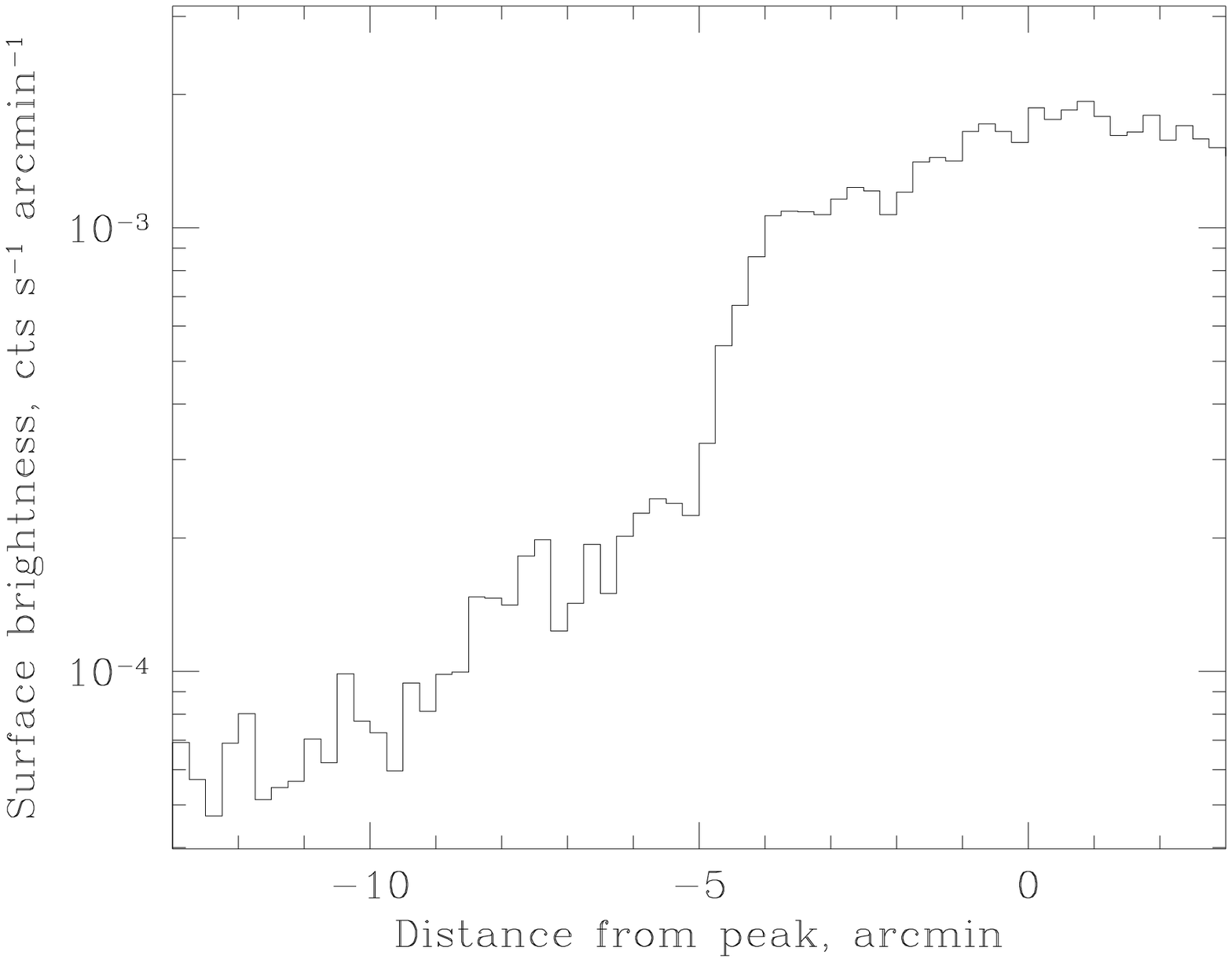}}

\rput[tl]{0}(-0.45,2.6){
\begin{minipage}{8.75cm}
\small\parindent=3.5mm
{\sc Fig.}~2.---A3667 \rosat\ PSPC linear brightness profile in a 3\am-wide
strip along the direction of the cluster elongation (from region 2 to region
8 in Fig.\ 1). A shock front is apparent at --5\am\ from the brightness
peak, inside region 2 of the temperature map. A similar feature is present
in the HRI image.
\end{minipage}
}

\endpspicture
%%%%%%%%%%%%%%%%%%%%%%%%%%%%%%%%%%%%%%%%%%%%%%%%%%%%%%%%%%%%%%%%%%%%%

The \rosat\ image suggests that one of these shock fronts has reached region
2 of our map (Fig.\ 2). We do not see an accompanying high temperature
there, while we do observe an elevated temperature at the collision site. It
is quite possible that measurement errors and projection effects have
reduced the temperature rise. There is also a more interesting but rather
speculative explanation: that there is a spatial lag between the gas density
jump and the corresponding electron temperature increase due to the lack of
equipartition between electrons and ions at the shock front and/or due to
the effects of thermal conduction (Shafranov 1957). It is unclear to what
extent electrons will be dissipatively heated in the moderate Mach number
collisionless shocks in cluster mergers. If the electrons are not heated
dissipatively, it takes about $t_{ep}\simeq 2\times 10^8\,{\rm
yr}\,\,(n_e/10^{-3}\,{\rm cm}^{-3})^{-1}\, (T_e/10^8{\rm K})^{3/2}$ for the
temperatures to equilibrate through electron-proton collisions (Spitzer
1962). During this time, a 1000\kms\ shock will travel $\sim 0.2$ Mpc.
Electrons will also be heated by adiabatic compression in the shock, and
this process is more important in weaker shocks, such as those in mergers.
For the crude estimate of the shock parameters given above, the compression
is $\rho_1/\rho_0 \approx 2.5$, consistent with the increase in X-ray
brightness in Fig.\ 2. This implies an adiabatic increase in the electron
temperature by a factor of $\sim$1.9.  It is possible that electron thermal
conduction, if not strongly suppressed, will act to smooth out this electron
temperature jump. Given the present crude temperature information and the
uncertainty of the merger stage, this discussion is very speculative.
However, future observations with \chandra\ might provide much more detailed
electron temperature and gas density profiles in merger shocks in A3667 and
other clusters. This would allow a direct study of electron heating, thermal
conduction (and therefore magnetic fields), and the rates of electron-ion
equipartition in the intracluster gas. In the meantime, our estimates
indicate the need to include these observable effects in the hydrodynamic
merger simulations. Indeed, simulations by Takizawa (1999) exhibit
qualitatively similar deviations of $T_e$ from the plasma mean temperature
at a late merger stage.

Assuming that, in general, $T_e$ represents the local mean temperature
reasonably well, it is interesting to note that specific entropy maps of
A3667 and Cygnus A (Fig.\ 1) show variations of the order of $\frac{3}{2}k$.
Energy that needs to be injected into the gas to produce such entropy
changes is $Q\simeq T\Delta s$. Therefore, such variations indicate that
energy input from the merger is comparable to the gas thermal energy,
$\frac{3}{2}kT$, providing direct empirical evidence that mergers at the
present time can contribute significantly to the heating of the intracluster
gas.

\section{A2065}
\label{sec:2065}

The temperature map of A2065 ($z=0.072$) reveals a hot region southeast of
the cluster peak which is probably a merger shock. The elongated galaxy
distribution in the inner $r\sim 10'$ (Postman, Geller, \& Huchra 1988) is
rather similar to the X-ray contours and does not show any distinct
subclusters. This suggests that the merger is at a late stage, perhaps well
after a core passage, although the data are too crude to discuss any
definitive scenarios. There is a marginally significant spectroscopic
indication of a mild cooling flow in the center (MFSV). Using a \rosat\ HRI
image, Peres et al.\ (1998) estimate a central cooling time of $\sim 4.4$
Gyr and $\dot{M}=13^{+14}_{-6}$\msunyr. Overlaying the \rosat\ HRI image
onto a Digitized Sky Survey plate (Fig.\ 3{\em a}) reveals two peaks in the
X-ray brightness, coincident with two central galaxies with a projected
separation of $\sim 35$ kpc (whose line of sight velocities differ by $\sim
600$ \kms; Postman et al.\ 1988). It is remarkable that these peaks have
survived a merger. We will use their survival to constrain the compactness
of the gravitational potential at the subcluster centers.

Many if not most clusters have sharp central gas density peaks due to the
cooling flows (e.g., Peres et al.\ 1998).  Because of the spectral
complexity of these regions, it is unclear if the gas pressure also is
peaked, as it would be if the dark matter distribution has a central cusp
(e.g., Navarro, Frenk, \& White 1995).  Alternatively, the central
temperature drop might keep the pressure gradient low, as required for a
dark matter distribution with a flat core such as a King profile. \asca\
spectral analyses of stationary cooling flows, although somewhat uncertain,
indicate the presence of a hot gas phase, supporting the first possibility
(e.g., Fukazawa et al.\ 1994; Ikebe et al.\ 1996), which is also consistent
with strong gravitational lensing results (Allen 1998). A merger in a
cooling flow cluster can provide an independent probe of the central gas
pressure, including its hypothetic nonthermal component (e.g., Loeb \& Mao
1994) unaccessible by the spectral analysis.

Whether the two central galaxies in A2065 are physically interacting or a
chance projection is not obvious from the optical data. If these galaxies
are at the centers of the colliding subclusters and are physically close
then both should have experienced the full effect of a merger sweeping the
collisional gas away from the collisionless galaxies and dark matter peaks
(while the latter would likely  survive; e.g., Gonz\'alez-Casado et
al.\ 1994). On the other hand, if they are separated along the line of sight
then it seems unlikely that both have avoided the shock passage. We assume
for definiteness that the shock in region 2 has passed over the southern
galaxy, although of course, this is not the only scenario.

This leads to two possibilities, that the observed gas density peak around
the galaxy (whether due to a cooling flow or not) has survived the shock, or
it has reestablished itself after the shock passage. Again neglecting the
uncertain merger geometry and stage for a qualitative estimate, from the
observed temperature variations $T_0\simeq 5$ keV and $T_1\simeq 14$ keV and
eqs.\ (1) and (2), one estimates $u_0-u_1\simeq 1800$\kms. Given this
velocity scale, the time since a shock may have passed the cluster center is
of order a few $\times 10^8$ yr, much shorter than the central cooling time
of 4.4 Gyr (Peres et al.\ 1998). Thus, if the hot gas has settled in the
gravitational potential after the shock passage, it should still be hot and
therefore, its peaked density distribution must reflect a peak of the
gravitational potential.

On the other hand, if this density peak has survived the shock passage
intact, then its gas pressure must be greater than the ram pressure of the
gas flowing around the peak, $p_{\rm peak}\gax p_{\rm ram}\sim \rho'
(u_0-u_1)^2$ (e.g., Daines \& Fabian 1991).  As a rough estimate of the
density of the gas exerting ram pressure, $\rho'$, we take the average
density of the central region of A2065, found by fitting a $\beta$-model to
the HRI brightness profile centered on the global cluster emission centroid,
which is slightly offset from the peak.  We use the average temperature of
5.5 keV and the cluster region inside 1.6 Mpc, simultaneously fitting the
constant HRI background. For this fit, we obtain $a_x^\prime=150$ kpc,
$\beta'=0.49$, $n_{H0}^\prime=5\times 10^{-3}$\cmcube\ and $\rho'=1.2\times
10^{-26}$\,g\cmcube, resulting in the estimate $p_{\rm peak}\gax 4\times
10^{-10}$\,erg\cmcube.  We can similarly estimate the gas density in the
peak, fitting a $\beta$-model to the brightness profile centered exactly on
the southern galaxy and using only the southern half of the cluster image to
exclude the brightness extension on the opposite cluster side. Although the
profile does not follow the $\beta$-model in the very center, it is still a
convenient representation for our purpose. The fit gives $a_x=50$ kpc,
$\beta=0.42$ and $n_{H0}=1.2\times 10^{-2}$\cmcube. With this $n_{H0}$, the
peak pressure estimated above corresponds to a temperature $T_{\rm peak}\gax
9$ keV, compared to the cluster average temperature of 5.5 keV.  The \asca\
measurement in region 1 of Fig.\ 3{\em b}, although uncertain, does not
exclude the existence of such a hot gas. Note, however, that the above high
pressure need not necessarily be of thermal nature.

We can now estimate the total mass around the southern galaxy using the
above pressure constraint.  For a rough estimate we assume that $p(r)\propto
\rho_{\rm gas}(r)$ (which corresponds to isothermal gas if the pressure is
thermal), that the gas is not far from hydrostatic equilibrium, and
substitute the above $\beta$-model parameters for the peak density
distribution to the hydrostatic equilibrium equation $dp/dr=-\rho_{\rm
gas}GM(r)/r^2$ (e.g., Sarazin 1988). For example, within $r=a_x=50$ kpc we
obtain $M\gax 1.1\times 10^{13}$\msun. For comparison, if one uses the
average cluster temperature of 5.5 keV and the gas density parameters of the
centroid fit $a_x^\prime$ and $\beta'$ obtained above, thereby assuming that
the density peaks on the galaxies are due solely to the isobaric cooling
flows, one obtains a 7 times lower mass within the same radius. Thus, the
survival of the gas density concentration centered on one of the giant
central galaxies indicates that there indeed is a sharp pressure peak at
that position, implying a sharply peaked total mass profile with the galaxy
at the peak.  Note that the estimated mass in the peak is too great to be
due to the galaxy itself ($M/L_B \ga 100\,M_\odot/L_\odot$, photometry from
Postman et al.\ 1988).

\section{SUMMARY}

Using \asca, \rosat\ and \einstein\ data, we derived gas temperature maps of
Cygnus A, A3667, and A2065. The maps indicate ongoing major mergers and
allow us to make several interesting qualitative estimates. (1) The observed
temperature variations in Cygnus A are close to those expected for a
free-fall collision of two subclusters with Cygnus A masses, suggesting that
the gas kinetic energy is effectively dissipated into heat during a merger.
(2) In A3667, we may be observing a spatial lag between the shock front seen
in the \rosat\ images and the rise of the electron temperature. A
considerable lag might be produced by the finite electron-ion temperature
equilibration time and the effects of thermal conduction.  (3) In A2065, the
double gas density peak centered on two central galaxies apparently has
survived a merger shock passage. From this, we infer a lower limit on the
total pressure in the peak which implies a strongly peaked total mass
distribution. (4) Specific entropy variations in A3667 and Cygnus A provide
direct empirical evidence that present-day major mergers contribute
significantly to the heating of the intracluster gas. Data from \chandra,
\xmm, and \astroe, together with improved simulations, will allow much deeper
insights into the physics of cluster mergers.

\acknowledgments

We thank Marc Postman for providing his A2065 optical image prior to
publication, and the referee for many useful comments. Support was provided
by NASA contracts NAS8-39073, NAG5-3057, NAG5-4516, and CfA postdoctoral
fellowship.

%%%%%%%%%%%%%%%%%%%%%%%%%%%%%%%%%%%%%%%%%%%%%%%%%%%%%%%%%%%%%%%%%%%
\begin{figure*}[tb]
\pspicture(0,6.5)(18.5,22.9)
%\psgrid(0,9)(18.5,23.0)

%%%%%%
\rput[tl]{0}(0.5,23.2){\epsfxsize=8.5cm
\epsffile{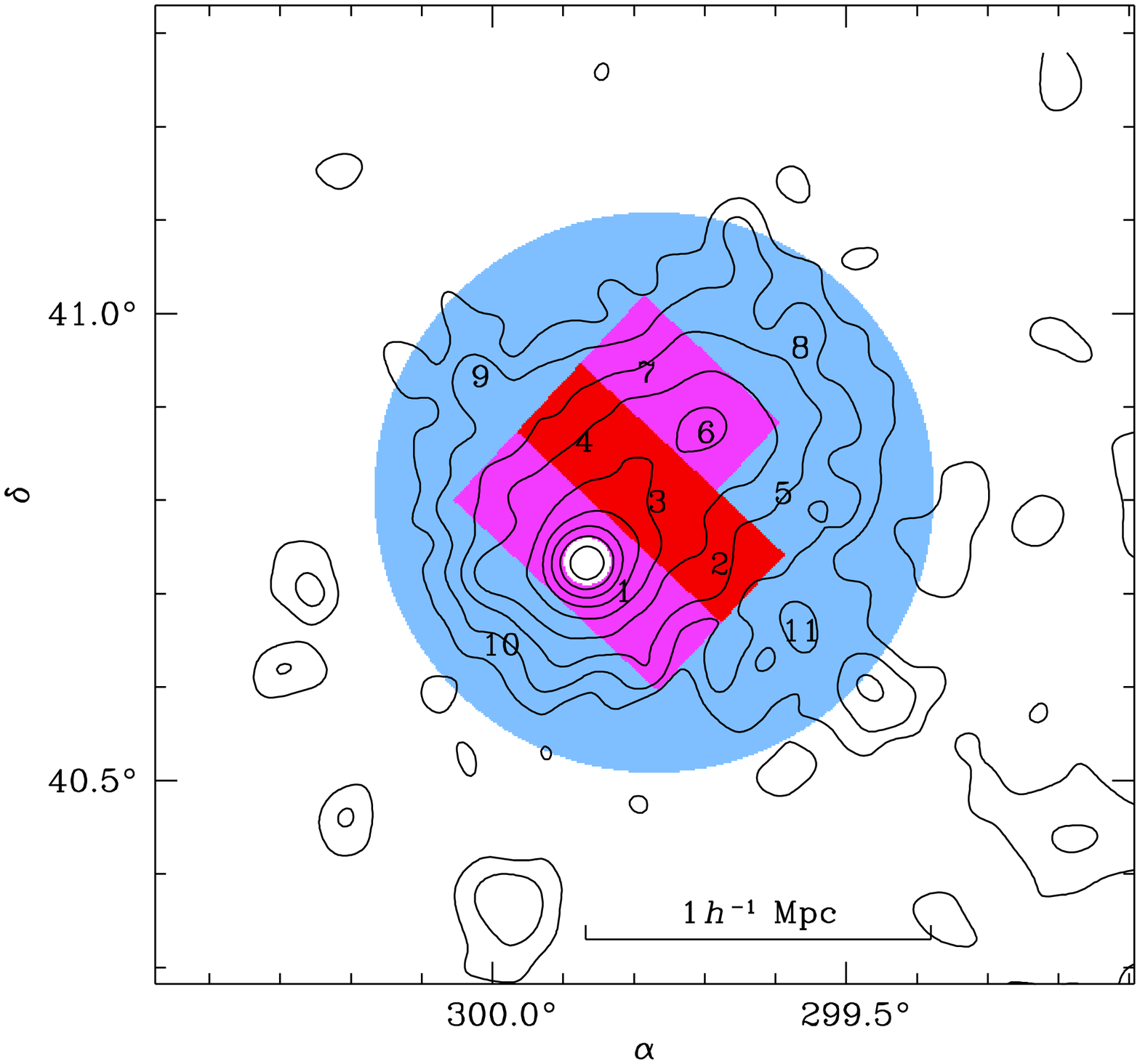}}

\rput[tl]{0}(0.75,16.2){\epsfxsize=7.75cm
\epsffile[30 470 530 678]{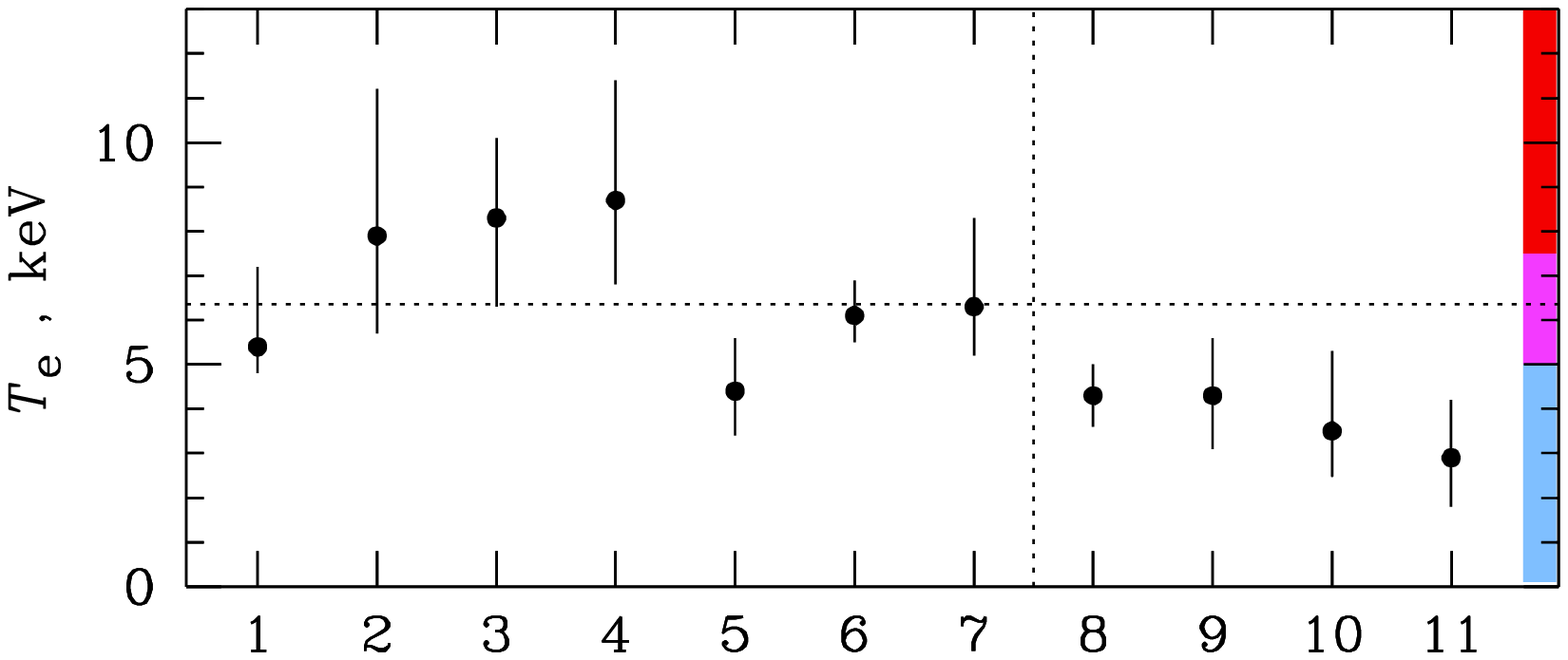}}

\rput[tl]{0}(0.75,13.5){\epsfxsize=7.75cm
\epsffile[30 428 530 678]{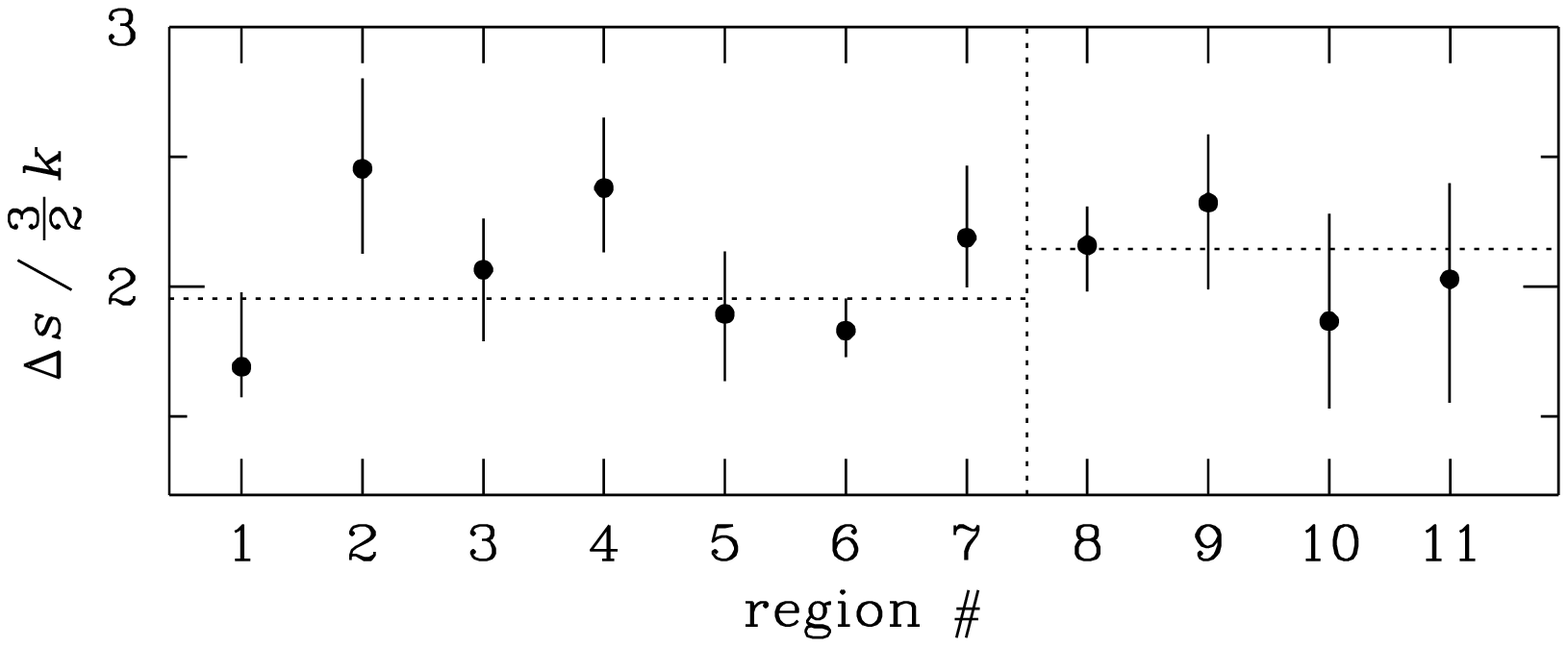}}

%%%%%%
\rput[tl]{0}(9.5,23.2){\epsfxsize=8.5cm
\epsffile{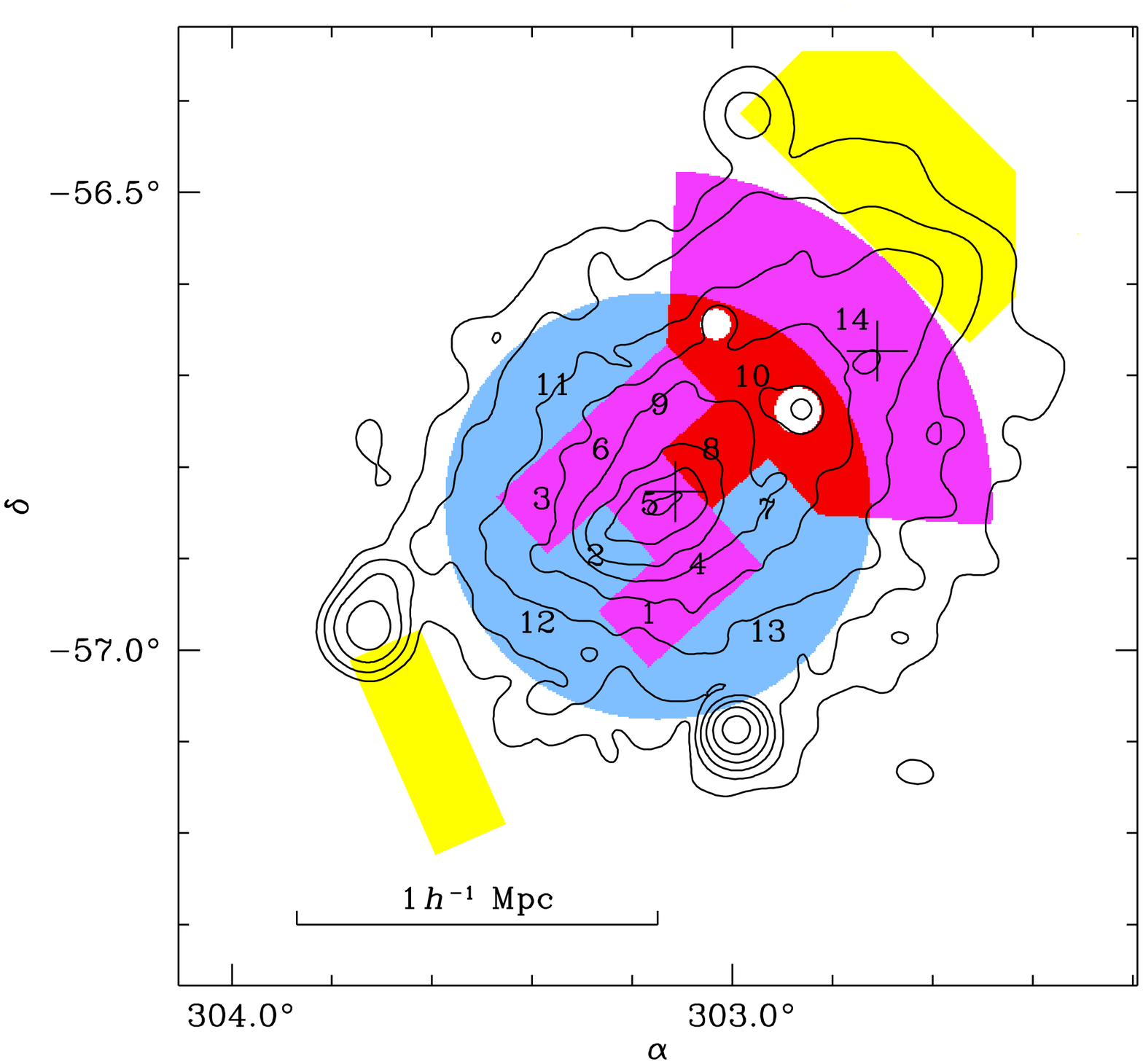}}

\rput[tl]{0}(9.75,16.2){\epsfxsize=7.75cm
\epsffile[30 470 530 678]{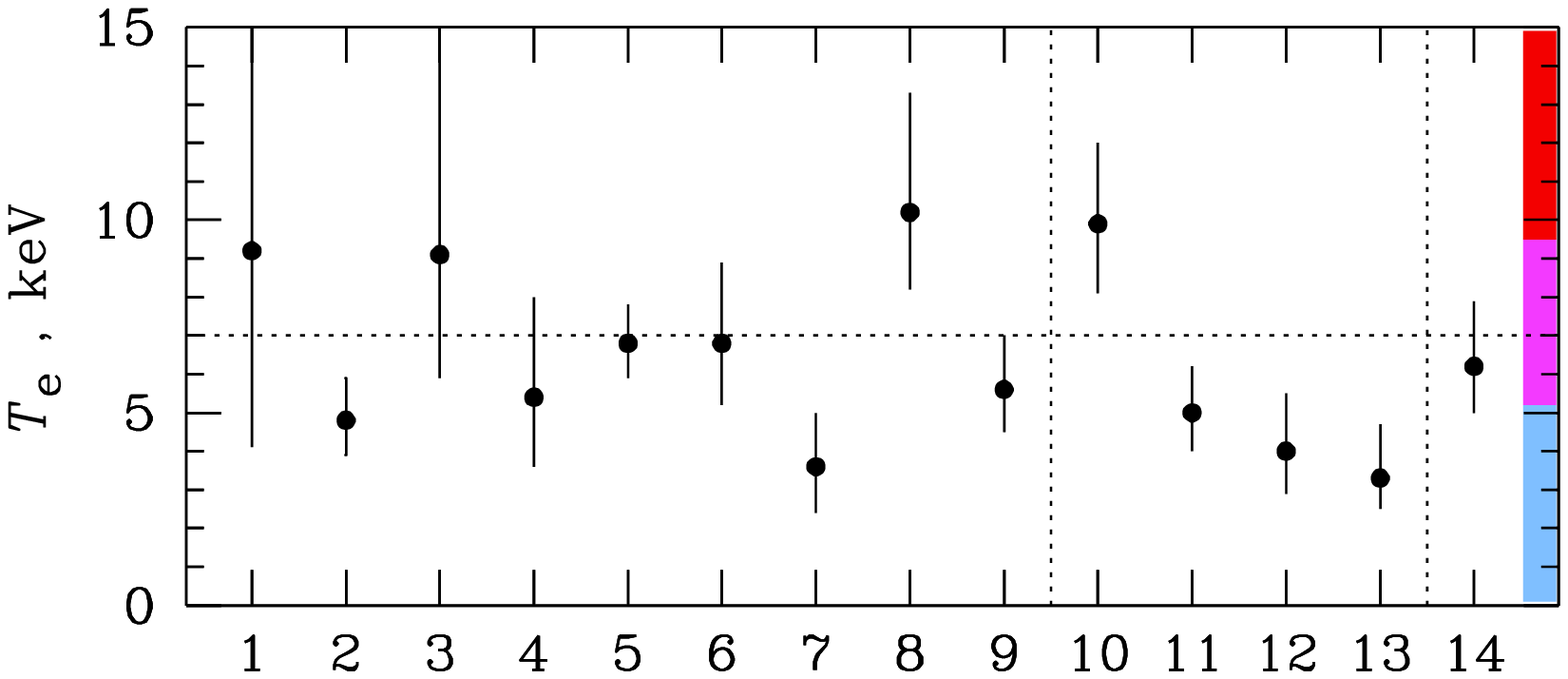}}

\rput[tl]{0}(9.75,13.5){\epsfxsize=7.75cm
\epsffile[30 428 530 678]{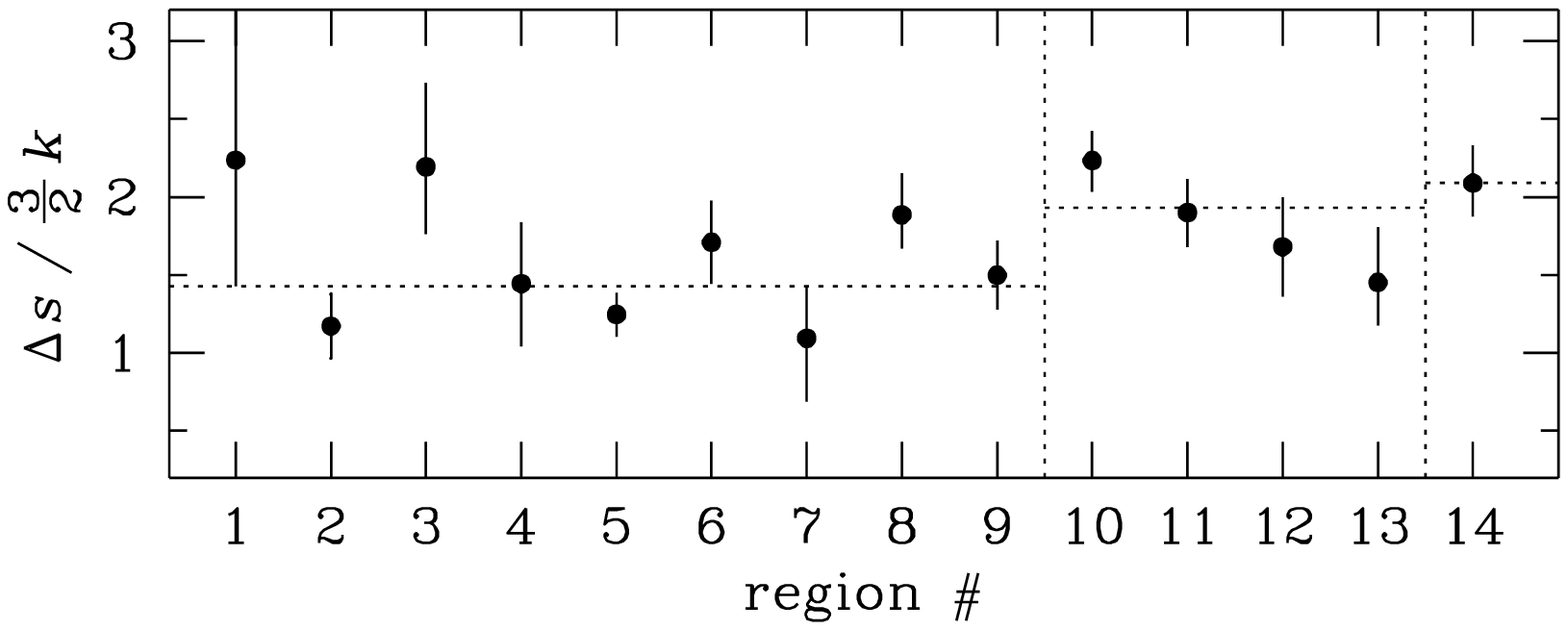}}

%%%%%%
\rput[l]{0}(2.3,22.5){\small Cygnus A}
\rput[l]{0}(11.3,22.5){\small A3667}

\rput[tl]{0}(0,9.3){
\begin{minipage}{18cm}
\small\parindent=3.5mm
{\sc Fig.}~1.---\asca\ projected temperature maps (color) overlaid on the
\rosat\ PSPC brightness contours spaced by a factor of 2. Regions
in which the temperature was derived are numbered in upper panels; their
temperatures and specific entropies with 90\% errors are given in lower
panels, along with the color scale for the temperature. Different colors
correspond to significantly different temperatures.  Dotted vertical lines
separate groups of regions belonging to the same annulus or a central
square.  Dotted horizontal lines show temperature averaged over the cluster
or entropy averaged over the respective annulus or square. White circles in
the maps show point sources either excluded or fitted separately (some are
not shown for clarity).  Region 1 in Cygnus A is a $6'\times 18'$ rectangle
with the AGN region excised. For A3667, crosses mark positions of the two
brightest galaxies, and yellow areas schematically show the radio halo from
R\"ottgering et al.\ 1997).
\end{minipage}
}
\endpspicture
\end{figure*}
%%%%%%%%%%%%%%%%%%%%%%%%%%%%%%%%%%%%%%%%%%%%%%%%%%%%%%%%%%%%%%%%%%%

%%%%%%%%%%%%%%%%%%%%%%%%%%%%%%%%%%%%%%%%%%%%%%%%%%%%%%%%%%%%%%%%%%%
\begin{figure*}[tb]
\pspicture(0,15.6)(18.5,23.1)
%\psgrid(0,0)(18.5,23.0)

\rput[tl]{0}(0.5,23.0){\epsfxsize=4.95cm \epsfclipon
\epsffile[47 148 550 651]{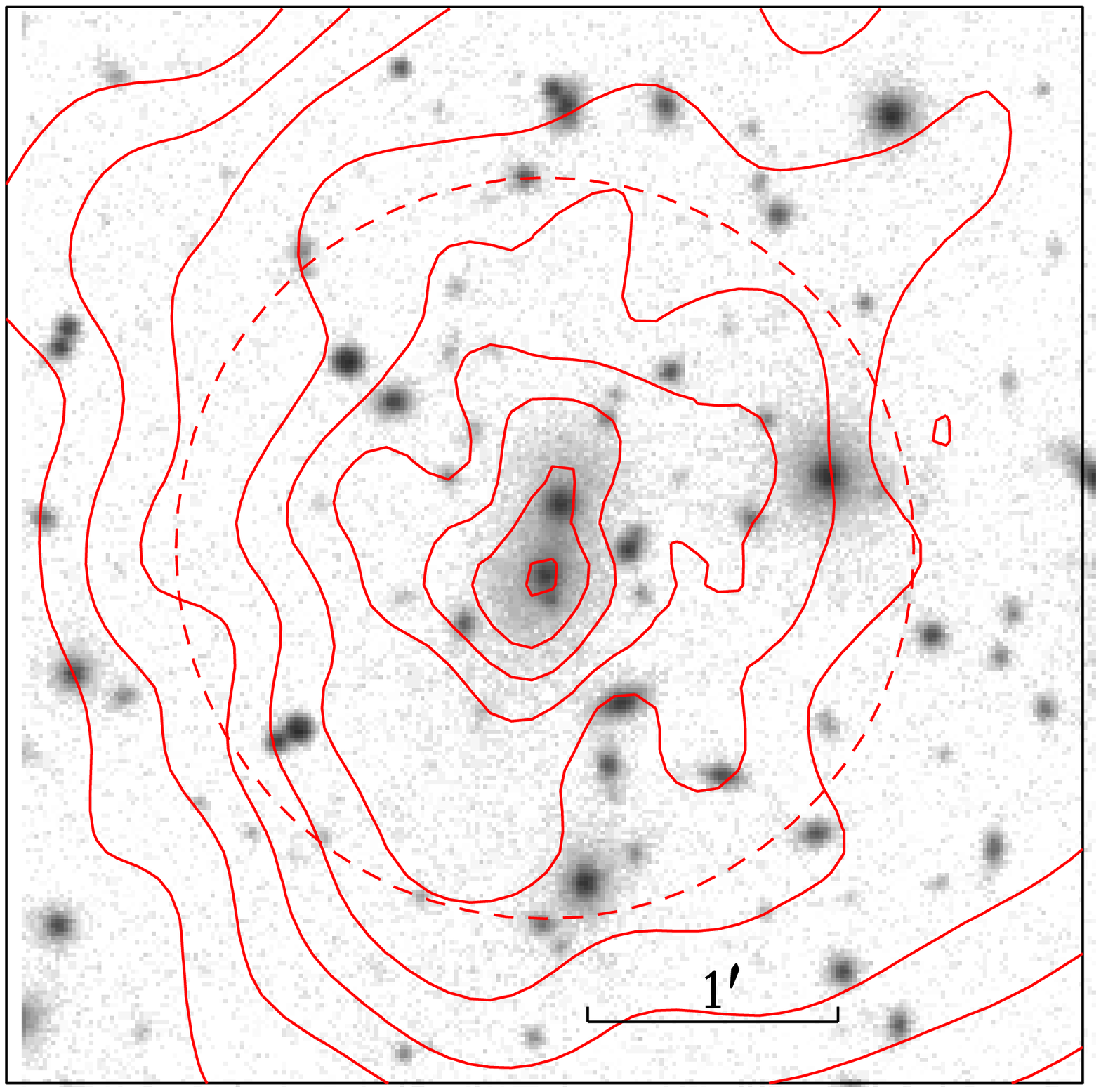}}

\rput[tl]{0}(5.9,23.15){\epsfxsize=6.5cm
\epsffile{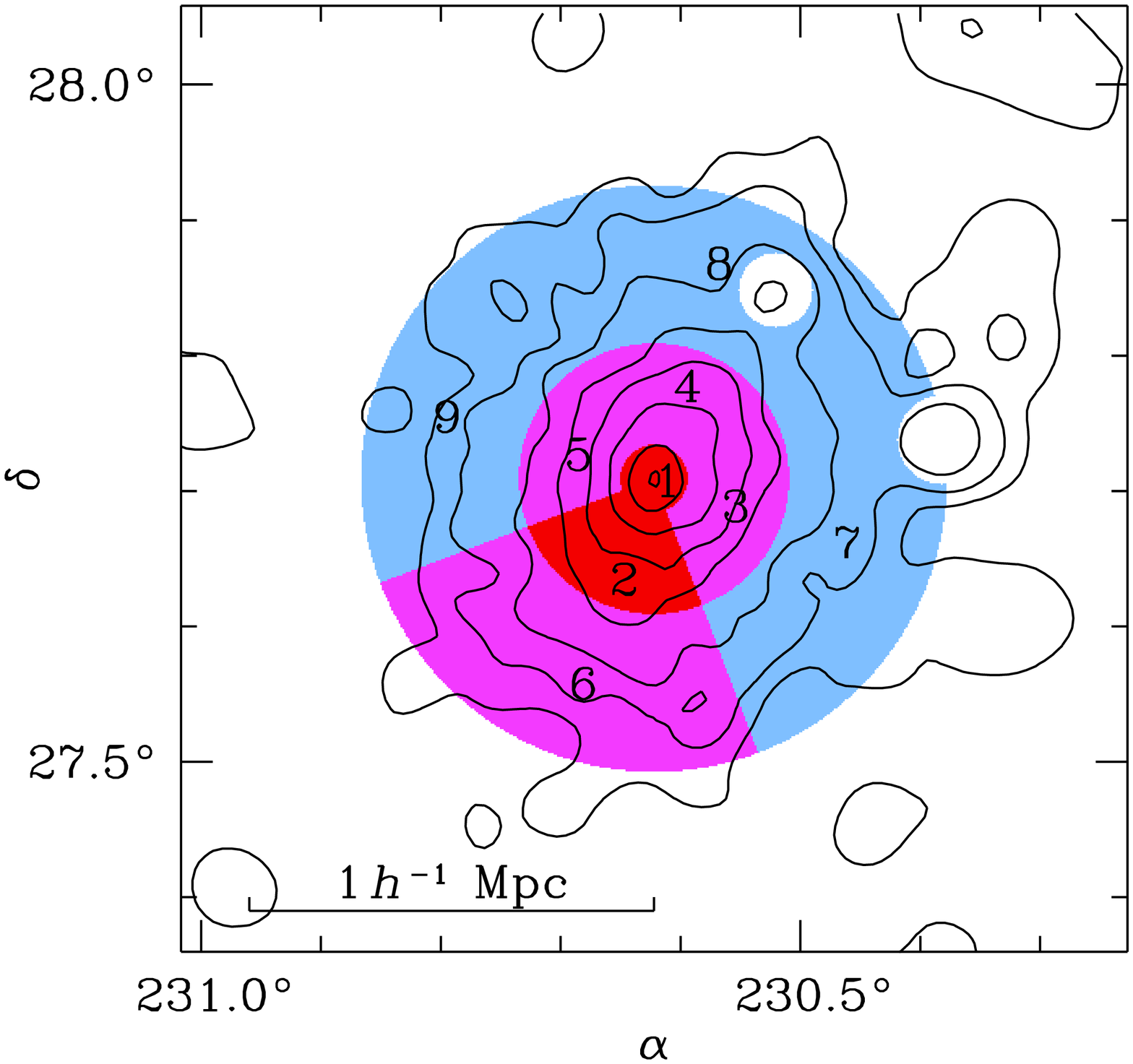}}

\rput[tl]{0}(12.0,22.7){\epsfxsize=8.7cm
\epsffile[30 428 530 678]{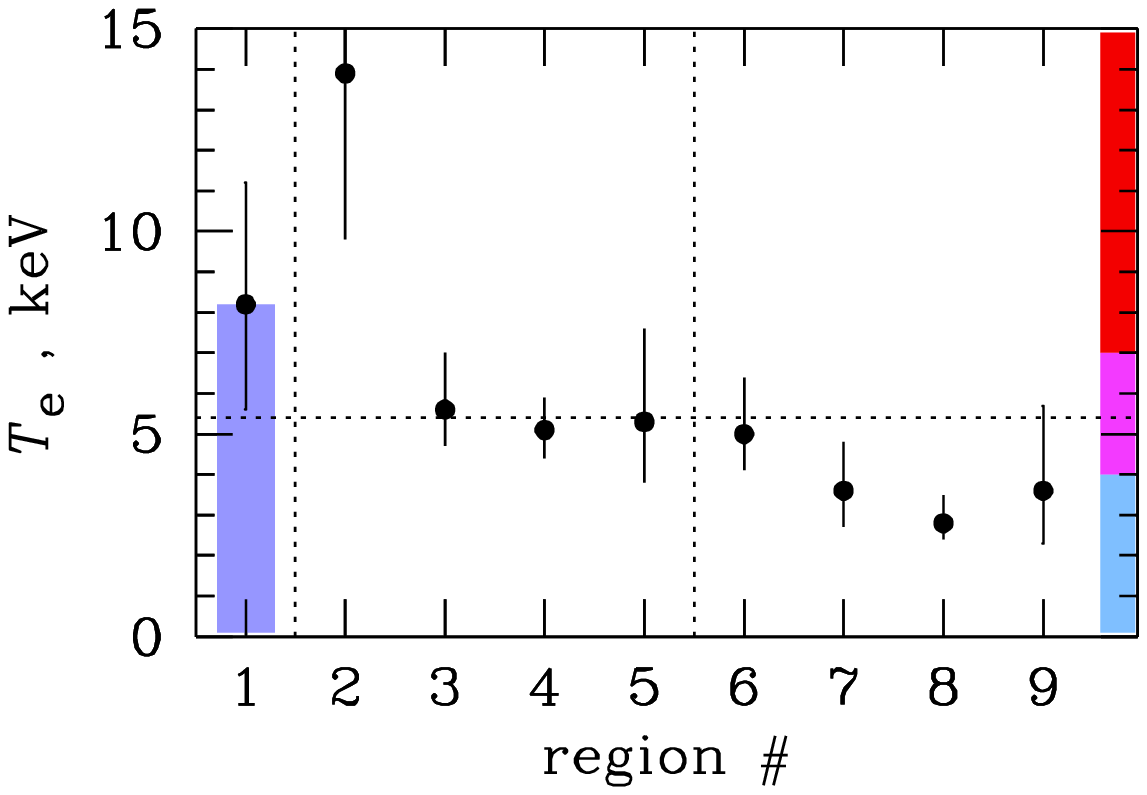}}

\rput[l]{0}( 0.7,22.7){\it a}
\rput[l]{0}( 7.2,22.7){\it b}
\rput[l]{0}(17.5,22.7){\it c}

\rput[tl]{0}(0,17.1){
\begin{minipage}{18cm}
\small\parindent=3.5mm
{\sc Fig.}~3.---({\em a}) \rosat\ HRI image of the central region of A2065
(contours) overlaid on the Digitized Sky Survey image. The contours are
spaced by a factor of $\sqrt 2$. An 8\as\ correction was applied to the HRI
coordinates using two point sources in the field of view. Dashed circle of
$r=1.5'$ corresponds to region 1 in panel ({\em b}), which shows
\asca\ temperature map (colors) overlaid on the \einstein\ IPC brightness
contours (spaced by a factor of 2). Regions are numbered and their
temperatures with 90\% errors are shown in panel ({\em c}).  For region 1,
ambient temperature of a cooling flow is shown and blue band denotes the
cooling flow.
\end{minipage}
}
\endpspicture
\end{figure*}
%%%%%%%%%%%%%%%%%%%%%%%%%%%%%%%%%%%%%%%%%%%%%%%%%%%%%%%%%%%%%%%%%%%

\end{document}